\documentclass[conference]{IEEEtran}
\IEEEoverridecommandlockouts
\usepackage{cite}
\usepackage{amsmath,amssymb,amsfonts}
\usepackage{graphicx}
\graphicspath{{figs/}}
\usepackage{textcomp}
\usepackage{booktabs}
\usepackage{array}
\usepackage{url}
\usepackage[a4paper, total={184mm,239mm}]{geometry}

\newcommand{\system}{ChipArc}

\begin{document}

\title{ChipArc: Auditable Multi-Agent RTL Generation, Repair and Optimization with Verifier-Gated Memory}

\author{\IEEEauthorblockN{Kaili Wang, Ben Stoffelen, Alejandra Lopez de Aberasturi Gomez, \\
Diederik M.\ Roijers, Ludovic Denoyer, Joyjit Kundu}
\IEEEauthorblockA{imec, AI-Labs
}}

\maketitle

\begin{abstract}
Register-transfer level (RTL) code must be functionally correct and efficient in
timing, area, and energy. We present \system{}, an auditable multi-agent framework that lets large
language models (LLMs) produce such code: models propose RTL and diagnose
failures, while a deterministic controller, not the model, checks every
candidate with the benchmark's native verifier, accepts it only if its rank
strictly improves, and logs each decision for offline audit. Agents on concurrent tasks share a memory of verified repair
records, excluding their own task's records and hidden test material. With Kimi K3, \system{}
solves 89 of 92 tasks of CVDP's non-commercial agentic code-generation subset at the lowest cost per pass of the compared systems, and 38 of 43 ChipBench
tasks on average, against 34 for ACE-RTL (rerun with the same model and one
trajectory per task instead of five) and 27 for MAGE. Starting from the same
unsolved designs for 33 CVDP tasks, the shared memory lets \system{} fix about
12\% more of them. With synthesis, bounded formal checking, and place-and-route
as verifiers, the same acceptance rule cuts a GCD/RSA accelerator's mapped area
by 68\%, latency by 74\%, and energy by 83\% relative to its LLM-generated
baseline, and its formal check exposes a protocol bug that simulation misses.
Finally, using reference designs (available for 216) and
testbench reruns with other seeds, we find 31 of our 448 CVDP designs
that pass the official testbenches to be functionally wrong; tests built from
the counterexamples reject 22 of them, and re-scoring with these tests leaves the system
ranking unchanged.
\end{abstract}


\section{Introduction}

Writing register-transfer level (RTL) code is a critical
step in chip design, because it determines what circuitry is built and therefore
how the chip behaves. RTL code describes a digital circuit's registers and
the logic that updates them, written manually in hardware-description languages such as
Verilog and mapped to gates by synthesis tools. Authoring and debugging RTL is
largely engineer-driven and costly; recent work uses large language models
(LLMs) to accelerate this work by drafting RTL and proposing repairs, evaluated on
benchmarks such as VerilogEval~\cite{verilogeval}, CVDP~\cite{cvdp}, and
ChipBench~\cite{chipbench} through repeated generate--verify--repair iterations.

A central difficulty in automatic iterative RTL repair is deciding whether an edit is actually an improvement. Model reasoning alone cannot establish that a candidate compiles, behaves correctly, or preserves functionality that already worked: even a one-line RTL edit can silently break previously correct behavior by changing signal timing, reset semantics, or
component interactions without triggering a syntax error. The system must therefore retain the best candidate under a deterministic ordering derived from the benchmark's verifier and accept an edit only when its verifier-measured progress strictly improves. A passing test suite alone does not close this gap: our physical-design case study exposes a protocol corner case by bounded formal checking, and Section~\ref{sec:formal} finds wrong designs that pass the official tests. Because hardware defects become expensive to correct later in the design process, engineers need both a passing result and a record of the evidence used
to accept each change.
Here, \emph{verified} means accepted by the configured verifier
(simulation tests on CVDP and ChipBench, plus bounded formal checking in
Section~\ref{sec:silicon}), not formal equivalence.

Prior work improves either the generated RTL---via domain-adapted,
RTL-specialized models~\cite{rtlcoder,codev,craftrtl}---or the repair process,
via agentic frameworks that add simulation feedback, reflection, sampling, and
generated testbenches~\cite{mage,verilogcoder,aivril,acertl}.
Several gaps remain. Existing agents do not demonstrate
a single auditable acceptance policy spanning functional verification,
synthesis, formal checking, and physical implementation
\cite{mage,verilogcoder,aivril,acertl}. In most systems the model chooses the next candidate \cite{verilogcoder,aivril,acertl}, although MAGE uses
simulator-derived ranking and rollback \cite{mage}. Verification can also be
expensive, particularly with sampled candidates or long iterative searches
\cite{mage,acertl}. Finally, these systems do not evaluate reuse of bounded, verified repair records across tasks that keeps where each record came from
\cite{mage,verilogcoder,aivril,acertl}, leaving negative transfer and
leakage of task information as open risks.
The closest systems cover parts of this problem. HORIZON~\cite{horizon}
uses one agent and a binary pass predicate, with no partial-progress ranking,
restart policy, or cross-task memory. Our KernelArc~\cite{kernelarc} shares
memory only among agents optimizing the \emph{same} kernel.

We propose \system{}, a multi-agent framework in which the LLM proposes RTL edits
and a deterministic controller decides, using the benchmark's native verifier,
whether to accept them. It keeps the best verified design, reverts regressions,
and checkpoints every transition for offline audit. Its individual mechanisms (keeping only the best verified candidate, restarts, shared
memory) are established; we claim their combination under one acceptance rule decided by the verifier and evaluate it component by component. Our contributions are:
(i) \emph{graded acceptance}: a deterministic rank over partial
verifier progress (Section~\ref{sec:architecture}) decides acceptance in repair, branches, and restarts, unlike a binary pass predicate~\cite{horizon}; (ii) \emph{cross-task
memory with leakage controls}: agents on \emph{different} concurrent tasks share
verified records, with same-task exclusion and sanitization, unlike same-task
memory~\cite{kernelarc}; (iii) \emph{model-free replay}: every acceptance decision can be re-checked from
stored artifacts, each identified by a digest of its content, without calling the model or the verifier
(tool and artifacts to be released);
(iv) a component-wise evaluation on CVDP and ChipBench, extended to an
RTL-to-layout flow; and (v) a study that checks designs passing the CVDP and ChipBench
testbenches against reference designs, a procedure that turns a confirmed
counterexample into a test for the benchmark's own harness, and a re-score of
all systems with these tests.

Finally, we apply \system{} to a complete RTL-to-layout optimization loop that applies the same verifier-controlled acceptance to synthesis, bounded formal proof, place-and-route, timing, and energy. On a GCD/RSA accelerator, starting from an LLM-generated baseline, the loop reduces mapped area by 67.6\%, latency by 74.0\%, and energy per transaction by 83.4\% relative to that baseline (not to an expert design), with timing closed at a 1.47ns clock period ($\approx$680MHz). Its bounded formal check also exposes a protocol bug that both the functional harness and a 24-test regression miss. The same architecture thus also applies to physical optimization.

\section{Related Work}

\subsection{RTL Generation and Repair}

Early systems translate natural language to Verilog and score it with executable
tests, from VeriGen and VerilogEval to the broader RTLLM and VerilogEval~v2
suites \cite{verigen,verilogeval,rtllm,verilogeval2}. Domain-adapted and
specialized models---ChipNeMo, RTLCoder, CodeV, CraftRTL, OriGen, and
ScaleRTL---improve generation via domain data, augmentation,
correct-by-construction synthesis, and compiler-guided self-reflection
\cite{chipnemo,rtlcoder,codev,craftrtl,origen,scalertl}, while RTLFixer,
VeriAssist, and RTL-Repair target syntax, self-correction, and symbolic repair
\cite{rtlfixer,veriassist,rtlrepair}. These methods improve the model or its
immediate feedback; we instead study how a deterministic controller directs a general LLM
during functional repair.

CVDP contains engineering-authored generation, modification, reuse, integration,
and debugging tasks \cite{cvdp}. We evaluate its complete 92-task v1.1.0
non-commercial agentic code-generation subset from categories cid003, cid004,
cid005, and cid016, and, in Section~\ref{subsec:heavy}, its 302-task
non-agentic and 75-task agentic-heavy splits. ChipBench broadens evaluation to realistic Verilog generation, debugging, and heterogeneous reference-model generation using designs drawn from open-source CPU IPs and competition platforms~\cite{chipbench}. We use the 43 tasks of its Verilog-generation split whose official verification assets are valid (Section~\ref{sec:eval}).

\subsection{Agentic Hardware Design}

VerilogCoder plans with a task-and-circuit graph and traces waveforms via an abstract-syntax-tree (AST)
tool \cite{verilogcoder}; MAGE uses multiple agents, 20 high-temperature
candidates, generated testbenches, and iterative editing \cite{mage}; and AIVRIL
couples generation with verification in the loop \cite{aivril}. ACE-RTL assigns
complementary roles to a fine-tuned RTL model and a general-purpose LLM: the
Generator writes and repairs RTL, the Reflector analyzes simulation feedback,
and the Coordinator carries that guidance into later iterations. It reports
an Agentic Pass Rate (APR) that counts a task as solved if any of five trials
solves it, each trial using five parallel processes and up to 30 iterations \cite{acertl}. HORIZON and RTL-Repo
extend agentic design to repository-level tasks \cite{horizon,rtlrepo}.
HORIZON reports single runs~\cite{horizon}, and its authors note that
a pass may only mean ``satisfies the visible harness''. Unlike HORIZON, \system{} ranks partial progress (Section~\ref{sec:architecture}).
\system{} also limits the number of model roles, uses each benchmark's native
verifier to decide acceptance, and records state transitions outside the LLM.

Cross-task memory can transfer interface, state-machine, and timing patterns but
can also fill the prompt with irrelevant guidance or expose
benchmark-specific information. KernelArc~\cite{kernelarc}, whose deterministic acceptance rule \system{}
inherits, shares memory among agents on the same kernel; HORIZON reuses its own
repository history~\cite{horizon}. Neither shares memory across \emph{different}
concurrent tasks, so neither needs to stop a task from seeing its own retries or
hidden assets through memory.
We store bounded records created after
verification rather than free-form summaries, exclude the current task, give each record a stable identifier, and keep where it came from.

\section{ChipArc Architecture}
\label{sec:architecture}

\begin{figure*}[t]
\centering
\includegraphics[width=0.92\textwidth]{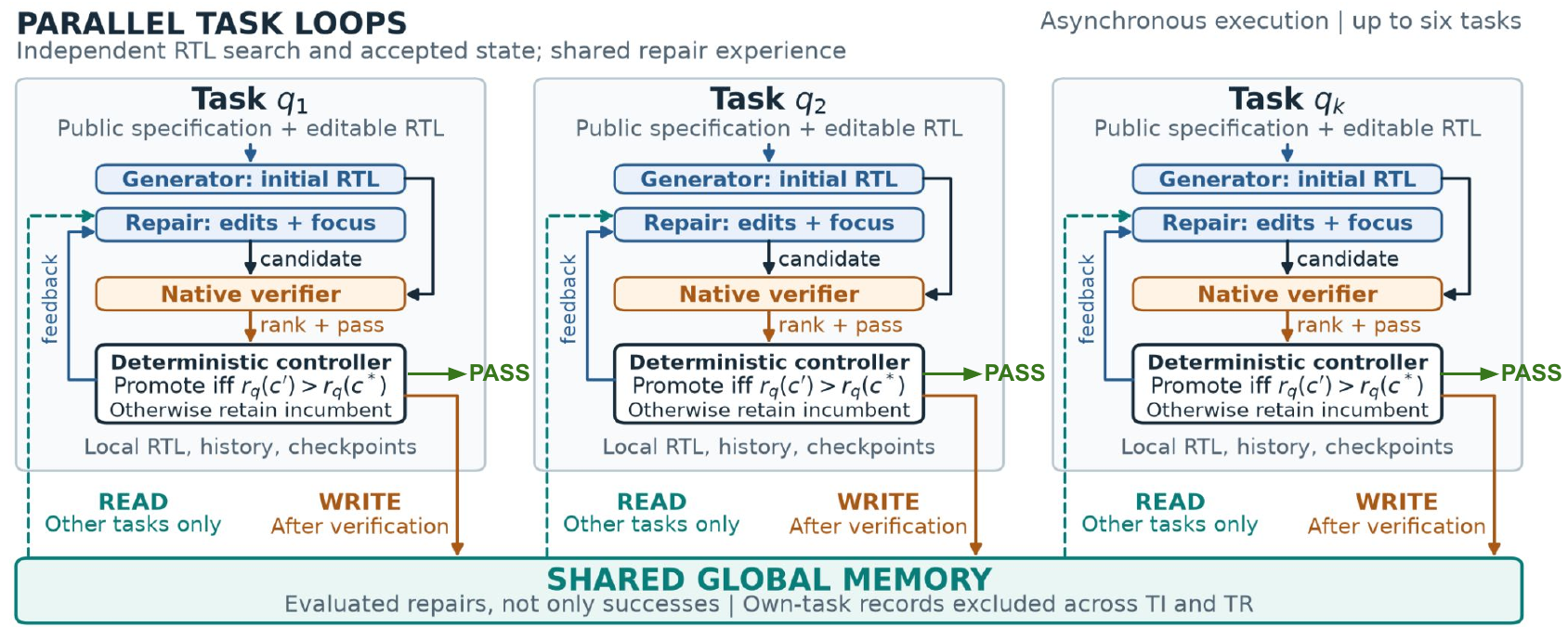}
\caption{\system{} runs independent task loops with one shared memory. Models
propose RTL; controllers verify and decide acceptance. Solid memory arrows write
sanitized records after verification; dashed arrows read only other-task records.
The initial (\textsc{ti}) and restart (\textsc{tr}) repair trails of a task retain separate local state and cannot share same-task records.}
\label{fig:system}
\end{figure*}

\system{} separates model proposals from verified state
(Figure~\ref{fig:system}). For each task, a \textbf{Generator Agent} drafts RTL
and a \textbf{Repair Agent} diagnoses failures and proposes edits; a
deterministic \textbf{controller} alone decides acceptance. Up to six task loops
execute asynchronously, sharing verified repair records through a bounded
global memory (GM)~\cite{kernelarc}. Content-addressed records make every acceptance decision auditable.

\subsection{Verifier-Controlled Acceptance}

For task $q$ (writing correct RTL given a specification), let $S_q$ be its public specification, $P_q$ its declared editable
paths, and $V_q(c)$ the native verifier result for candidate workspace $c$. The
benchmark adapter converts $V_q(c)$ into a pass bit and a deterministic rank
$r_q(c)$, a tuple of verifier measurements compared left to right and only
within the same task. On CVDP, candidates are ordered first by how far
they get (infrastructure error, timeout, compilation, elaboration, harness
failure, invalid runtime values, assertion failure, pass). Ties are broken by test
progress (pass ratio, then failed tests and milestones, mismatches, and failed
services) and finally by the smaller numeric error. On ChipBench, they
are ordered by whether they compile, finish in time and report a result, then by
the fraction of samples that match the reference and by how late the first
mismatch occurs. Model self-assessment and code size never enter $r_q$.

The controller maintains one accepted candidate $c_t^*$, called the incumbent.
The controller accepts (\emph{promotes}) a proposal $\tilde{c}_t$ only if its rank strictly improves:
\begin{equation}
c_{t+1}^*=\begin{cases}
\tilde{c}_t, & r_q(\tilde{c}_t)>r_q(c_t^*),\\
c_t^*, & \text{otherwise}.
\end{cases}
\label{eq:promotion}
\end{equation}
The first passing candidate is accepted and terminates the task. The LLM neither
computes the rank nor declares a pass: the guarantee is \emph{rank monotonicity}
within each repair trail under the configured verifier, not correctness beyond
what it checks. Hidden reference RTL, private testbench source, and golden
patches remain verifier-only; agents see public task material, candidate RTL,
and size-limited verifier output.

\subsection{Bounded Repair and Exploration}
\label{subsec:repair}

The Generator receives $S_q$, $P_q$, public repository context, and target file
contents. It emits up to five independently verified candidates as complete
files or, on CVDP, as replacements of text spans that must occur exactly once in the file. Unless one passes, the
highest-ranked candidate seeds the initial repair trail (\textsc{ti}); this candidate is the first incumbent $c_0^*$, with ties going to the
earliest generated candidate. Generator
history and discarded branches are not transferred. In the ablations of Section~\ref{subsec:components}, which start every run from the same frozen set of failed candidates,
\textsc{ti} instead starts from that supplied failed candidate.

Each Repair Agent turn receives the current candidate, structured verifier
evidence, accept or revert decisions, local memory of accepted
candidates, bounded conversation history, and GM records from other tasks. It returns
a structured \emph{focus} (failure, hypothesis, evidence, strategy, expected
effects, and abandonment conditions), a request to continue or roll back, and proposed files or edits. Before verification, the controller rejects
malformed responses, undeclared paths, ambiguous edits, and invalid workspace
updates. Benchmark adapters supply bounded, deterministically ordered evidence:
diagnostics, test outcomes, mismatch information, and waveform windows, each
tagged with its source and visibility.
To explore edits that do not immediately improve the verifier rank, the controller may continue from a runnable rejected candidate in a temporary branch after a revert. On ChipBench this requires that the agent asks to continue, that the candidate
compiles and does not time out, and that its mismatches do not increase. On CVDP the candidate must keep all previously passing tests and
reach a functional or structural stage; it is admitted when the agent asks to continue or when a deterministic check detects progress on the failure
currently being repaired. Each branch is limited to depth three and six repair rounds. Branches cannot replace
the incumbent without satisfying Equation~\eqref{eq:promotion}.

\textsc{ti} runs until a candidate passes or its round, stagnation, or cost limit is
reached (on CVDP and ChipBench, respectively, 18 and 16 repair rounds and a
cost cap of \$11 and \$4.50 per task; on both, a stagnation window of 10). If unsuccessful, a fresh restart trail (\textsc{tr}) repeats best-of-five generation
and repair with an independent workspace, history, and local memory. \textsc{ti} and \textsc{tr}
never merge candidates or exchange same-task records, including through GM.
If both attempts are exhausted, the task is marked failed and the final
incumbent is returned, never an unaccepted proposal.

\subsection{Verifier-Gated Cross-Task Memory}
\label{subsec:gm}

GM shares repair knowledge across concurrent task loops, such as interface
conventions, reset patterns, and failed repair strategies. It targets batches of related tasks, such as a benchmark suite or one
IP family, that share interface conventions, coding style, and harnesses. Only after native
verification does a controller commit a bounded record
$m_i=(f_i,a_i,\Delta_i,v_i)$: sanitized focus, requested action, change summary,
and verifier outcome. The hypothesis and measured outcome remain separate:
evaluation neither certifies the hypothesis as causal nor implies the repair
succeeded. The change summary contains anonymous target metadata, edit counts,
and bounded diff excerpts
(at most four files, three hunks per file, two lines per side); complete RTL and full diffs remain local. Deterministic digests bind
records to the dataset, task ordering, schema, protocol, and context limit.
Each campaign (one repetition of one model on one benchmark) starts from an
empty GM, and a campaign fingerprint prevents reuse of its records. Records thus
never carry over between repetitions, models, or benchmarks (including each
repetition of the 33-task ablation) and accumulate only within one campaign (139--304 records
in our runs). Each repair turn logs the snapshot it saw, so replay does not
depend on scheduling.

Before each repair turn, the controller retrieves only other-task records,
$\mathcal{M}_{-q}=\{m_i\mid\operatorname{task}(m_i)\neq q\}$. Every such record appears as a compact entry in a fixed order. Entries are grouped by
source trace, with no ranking or omission; long fields are clipped, and a turn
fails rather than silently dropping records if the index (the list of entries) exceeds 1M characters.
The agent may cite handles to request bounded details for at most
three records on the next turn; the current specification and verifier evidence
always take precedence. Same-task exclusion applies across both \textsc{ti} and \textsc{tr},
making this cross-task transfer rather than communication between retries.
Sanitization removes identifying paths, task IDs, RTL filenames, test names,
long literals, explicit expected/actual fields, hidden-source references, and
instruction-like text. Deterministic regular expressions sanitize each record before it is
committed; signal names are kept, and records are not checked against
hidden-asset fingerprints. A typical record states that the harness found no
output port, that a minimal one-file edit added it (18 lines added, 4 removed),
and that the status rose from runtime error to assertion failure, so the edit
was accepted.

\subsection{Auditability}
\label{subsec:audit}
For each evaluation, the controller logs the candidate identifier, verifier rank and pass status, cost, and cumulative verifier calls. Each repair record links digests of the baseline and edited candidate to the agent's focus and action, before/after ranks, and acceptance decision. Candidates and evidence are stored by content digest, while configuration and dataset manifests support model-free replay of acceptance decisions. After each transition, an atomic checkpoint saves the incumbent, verifier result, resource counters, local memory, and search state, allowing execution to resume from the last verified state.
For replay, the audit tool reads only stored artifacts, querying
neither model nor verifier. It checks that each acceptance obeys
Equation~\eqref{eq:promotion} against the recorded incumbent rank, that no branch
overwrites the incumbent, that final checkpoint digests and ranks match, and that
every GM snapshot can be rebuilt with same-task exclusion. On the Kimi K3 CVDP
and ChipBench campaigns, it reproduced all 345 final checkpoints and 866 acceptance decisions without violation. The released audit tool and run artifacts will include a replayable trace of
CVDP task DES 0007, a Data Encryption Standard (DES) design: starting from a failing candidate
(pass ratio 0.67), the agent makes two reverted proposals and then a passing one
that cites a GM record from another DES task (eight verifier calls, \$1.39).

\section{Experimental Evaluation}
\label{sec:eval}

\subsection{Research Questions}

Our experiments address five research questions.
\textbf{RQ1 (overall):} how does
\system{} compare with recent publicly available agentic RTL
systems in task completion, model cost, and native-verifier use on CVDP and
ChipBench? \textbf{RQ2 (per-task search):} under a single repair attempt, does
\system{}'s verifier-controlled repair outperform an agentic baseline, and how
much does the fresh restart contribute on matched CVDP
tasks? \textbf{RQ3 (cross-task memory):} how does the shared global memory
change repair coverage, native-verifier use, and model cost under the same
matched conditions? \textbf{RQ4 (physical optimization):} can the architecture
extend to RTL-to-layout optimization while satisfying the configured simulation
and bounded formal checks, and how do measured area, timing, and energy change
relative to the baseline within a given budget? \textbf{RQ5 (test adequacy):} how often are designs that
pass the official testbenches functionally wrong, and how do scores change
under strengthened tests?

\subsection{Baselines}

To compare \emph{control policies} rather than the underlying models, we evaluate
two agentic RTL baselines rerun with the same model as \system{} (Kimi K3 or
DeepSeek V3.2) and comparable budgets. MAGE samples 20 (by default high-temperature) initial
candidates, generates its own testbenches, and iteratively edits the best
candidate \cite{mage}. ACE-RTL combines an RTL-specialized generator---originally
a Qwen model trained on 1.7M RTL samples---with a reasoning LLM through
generator, reflector, and coordinator components, running five trials of five
parallel processes and up to 30 iterations \cite{acertl}. We normalize two things
for fairness: we replace ACE-RTL's specialized generator with the shared model,
and we run it \emph{serially}, one trajectory per task instead of five in parallel, to match
budget; MAGE's 20 candidates per task are left unchanged.
Both baselines otherwise retain their original search policies.
ACE-RTL is thus a \emph{modified}
configuration that tests its control policy with our model; the published system
(fine-tuned generator with Claude~4), whose generator is not released, may
perform better. All systems run six tasks concurrently. MAGE ranks candidates with the official
verifier, as \system{} does; in our runs it samples at the provider's default temperature. All systems
share the model, reasoning setting, verifier, and 900\,s CVDP timeout; ACE-RTL
and \system{} also share the per-task cost cap (\$11 on CVDP, \$4.50 on
ChipBench), which MAGE has on CVDP only. These caps bound but do not equalize
spend or verifier calls, so the comparisons are not compute-matched.

\subsection{Task Inputs, Verifiers, and Metrics}

We give every system the same model-visible inputs. The Generator sees the
specification and public context: for CVDP, the mini-repository, declared
editable paths, and current target contents (the non-commercial dataset file marks
target files with empty patches and contains no solutions); for ChipBench,
the required top-module interface and any embedded helper RTL.
Evaluator-only assets never appear in model prompts: the original system message, the private harness, the CVDP test harness (Docker, cocotb, pytest), and ChipBench's testbench and golden reference module.

Both verifiers measure functional test-suite completion, not formal equivalence:
ChipBench requires zero mismatches against the hidden golden reference module,
and CVDP requires all services of the official Docker/cocotb/pytest harness to
complete. CVDP spans 34 specification-to-RTL, 25 code-modification, 22
reuse/integration, and 11 debugging tasks (17 easy, 55 medium, 20 hard); of
ChipBench's 45 Verilog-generation tasks, 43 remain after a pre-run verifier check excludes one
task whose reference and checker disagree and one with a testbench syntax error.

As metrics, \emph{Final} is the number of tasks that pass the official
verifier, reported out of the task count (e.g., 92, 43, or 33). \emph{Cost} is total model
spending in USD, \emph{Cost/pass} is cost per solved task, \emph{Evals} is the
number of native-verifier invocations (each runs the full simulator or harness and
accounts for most of the non-model compute), and \emph{Evals/pass} is verifier
runs per solved task. Higher \emph{Final} is
better; lower cost and evaluation counts are better. $n$ denotes independent repetitions per row (mean $\pm$ sample
std). For significance, each system's per-task solve rate is averaged over its
repetitions, and differences are tested over tasks with a paired bootstrap (95\%
confidence interval, CI) and an exact sign test on tasks whose rates differ. The component study also
reports \emph{\textsc{ti} solved} (tasks solved in the initial repair trail, \textsc{ti}),
\emph{\textsc{tr} tried/solved} (the subsequent fresh-restart trail, \textsc{tr}), and
\emph{Cost/pass@\textsc{ti}} (\textsc{ti} cost per \textsc{ti} pass). For the
full 92-task CVDP comparison (Table~\ref{tab:cvdp-main}), all local campaigns use
the same dataset snapshot, Kimi K3 with medium reasoning, the official Docker
verifier%
, and six concurrent workers. They differ only in
how they organize model and verifier calls. For ChipBench, ACE-RTL and \system{}
use the same 43-task set, model, and checker in each panel, with three repetitions
per system.

\subsection{Overall Results (RQ1)}

\begin{table}[t]
\caption{CVDP results on 92 tasks with Kimi K3. Panel~1: full-92 campaigns
($n=3$); panel~2: post-hoc repair of the 33 tasks a shared Generator stage failed (frozen-33, Section~\ref{subsec:components}; $n=1$). Bold marks the best
per panel. $n$: independent repetitions. ACE-RTL: modified serial configuration;
local: rerun by us (Section~\ref{sec:eval}).}
\label{tab:cvdp-main}
\centering
\footnotesize
\setlength{\tabcolsep}{3pt}
\resizebox{\columnwidth}{!}{%
\begin{tabular}{l@{\hspace{2pt}}rrrrr}
\toprule
\multicolumn{6}{l}{\textit{Full-92 end-to-end campaigns}} \\
\multicolumn{3}{l}{System} & Final (/92) & Cost/pass & Evals/pass \\
\midrule
\multicolumn{3}{l}{MAGE local} & $80.7\!\pm\!4.2$ & $1.4\!\pm\!0.0$ & $14.4\!\pm\!1.2$ \\
\multicolumn{3}{l}{ACE-RTL serial} & $83.3\!\pm\!1.5$ & $1.5\!\pm\!0.2$ & $\mathbf{5.0\!\pm\!0.6}$ \\
\multicolumn{3}{l}{\system{}} & $\mathbf{88.7\!\pm\!0.6}$ & $\mathbf{1.2\!\pm\!0.2}$ & $5.3\!\pm\!0.7$ \\
\midrule
\multicolumn{6}{l}{\textit{Post-hoc local comparisons}} \\
System & Shared initial & Repairs (/33) & Final (/92) & Cost/pass & Evals/pass \\
\midrule
ACE-RTL serial & 59 & 24 & 83 & 1.3 & 4.9 \\
\system{} & 59 & \textbf{32} & \textbf{91} & \textbf{1.1} & \textbf{4.8} \\
\bottomrule
\end{tabular}}
\end{table}

On the full 92-task CVDP benchmark (Table~\ref{tab:cvdp-main}), \system{} solves 89, 89, and 88 tasks
($88.7\pm0.6$), always failing Digital Stopwatch and Phase Rotation 0031/0038
(the third run also misses the HDBn codec).
ACE-RTL solves 83, 82, and 85 tasks ($83.3\pm1.5$), while MAGE solves 76, 84,
and 82 ($80.7\pm4.2$). \system{} also has the lowest mean total model cost
($\$105.8$, compared with $\$125.7$ for ACE-RTL and $\$115.6$ for MAGE) and
the lowest cost per solved task ($\$1.19$/pass). Its mean verifier count is
comparable to serial ACE-RTL (468.7 versus 415.7), whereas MAGE uses 1,164
evaluations, about 2.5 times as many.  The gain is significant: $+5.3$ tasks over ACE-RTL (95\% CI $[1.0,10.0]$;
better on 10 tasks, worse on 1, sign test $p=0.012$) and $+8.0$ over MAGE
(CI $[4.0,12.7]$; 14 vs.\ 0, $p<0.001$).
Two factors explain these results. MAGE spends most verifier calls on its 20 sampled
candidates per task before repair. ACE-RTL and \system{} use similar verifier
budgets but spend them differently: on CVDP, ACE-RTL always continues from its
latest candidate, so a regressing edit becomes the next base. \system{}
instead never builds on a lower-ranked candidate, except in temporary bounded branches, and restarts
after stagnation. \system{} also needs fewer model calls per solved task (about 6,
versus 13 for ACE-RTL and 15 for MAGE). All of \system{}'s misses are assertion failures. On Phase Rotation 0031 and 0038, final candidates pass 0/50 and 30/60 tests, and no baseline solves either. On Digital Stopwatch, \system{} passes at most 1/4 tests, whereas
ACE-RTL solves it in all runs. \system{} solves Digital Stopwatch and Phase Rotation 0038 in two repetitions of the frozen-33 repair study (Section~\ref{subsec:components}), so these misses
partly reflect the Generator's starting candidates.

\begin{table}[t]
\caption{ChipBench results on 43 verifier-valid tasks, grouped by model. ACE-RTL
and \system{} rows are mean $\pm$ std over three runs ($n=3$); MAGE published is
a single reported value on the 45-task set (17/45) and is not directly comparable. ACE-RTL:
modified serial configuration. Not compute-matched.}
\label{tab:chipbench}
\centering
\small
\setlength{\tabcolsep}{3pt}
\resizebox{\columnwidth}{!}{%
\begin{tabular}{llrrr}
\toprule
Model & System & Final (/43) & Cost/pass & Evals/pass \\
\midrule
Kimi K3 & MAGE local & $27.0\!\pm\!1.0$ &
$0.8\!\pm\!0.1$ & $22.7\!\pm\!2.2$ \\
& ACE-RTL serial & $34.3\!\pm\!1.5$ &
$\mathbf{0.7\!\pm\!0.1}$ & $\mathbf{7.4\!\pm\!0.4}$ \\
& \system{} & $\mathbf{38.3\!\pm\!0.6}$ &
$0.8\!\pm\!0.1$ & $7.7\!\pm\!0.5$ \\
\midrule
DeepSeek V3.2 & MAGE published & 17 & -- & -- \\
& ACE-RTL serial & $29.3\!\pm\!2.3$ &
$\mathbf{0.04\!\pm\!0.01}$ & $\mathbf{12.0\!\pm\!2.7}$ \\
& \system{} & $\mathbf{33.7\!\pm\!1.2}$ &
$0.16\!\pm\!0.03$ & $15.1\!\pm\!1.1$ \\
\bottomrule
\end{tabular}}
\end{table}

As Table~\ref{tab:chipbench} shows, with Kimi K3, \system{} solves a mean of
$38.3\pm0.6$ tasks, compared with $34.3\pm1.5$ for serial ACE-RTL and
$27.0\pm1.0$ for MAGE. Relative to ACE-RTL, this is a mean gain of 4.0 tasks
at 24.1\% higher model total cost and 16.6\% more total evaluations. On a
per-solved-task basis the gap is smaller: \system{} costs \$0.8/pass versus
\$0.7 for ACE-RTL and \$0.8 for MAGE, so its higher total spend mainly
reflects the extra tasks it solves. MAGE solves the fewest tasks and uses the
most verifier evaluations ($613.3\pm82.0$). With DeepSeek V3.2, \system{} solves
$33.7\pm1.2$ tasks, compared with $29.3\pm2.3$ for ACE-RTL, a mean gain of 4.3 tasks at 4.3$\times$ the model cost and 45.6\% more evaluations. Its cost
per solved task is \$0.16 versus \$0.04 for ACE-RTL. \system{} therefore solves
more tasks with both models, while serial ACE-RTL uses fewer verifier calls.
Neither comparison is compute-matched.
With DeepSeek V3.2 this trade-off is unfavorable, and its cause is input size per call: \system{} sends
about 45k prompt tokens per call over roughly 500 calls per run, versus 4k over
800--1{,}100 calls for ACE-RTL. The full GM index accounts for an estimated two
thirds of repair prompt tokens, and \textsc{tr} consumes 46\% of spend for 4 of
101 passes (Kimi K3: 28\%, 3 of 115). Compacting the index and starting \textsc{tr} only when it is likely to help could reduce this cost.

\begin{table*}[t]
\caption{Matched frozen-33 component results (33 tasks failed by a shared Generator stage), mean $\pm$ std over three
runs ($n=3$). \textsc{tr} reports tried/solved; Long-\textsc{ti} continues the failed \textsc{ti} for the combined \textsc{ti}+\textsc{tr} budget instead of restarting. Bold marks the
best of the first two rows.}
\label{tab:components}
\centering
\scriptsize
\setlength{\tabcolsep}{5pt}
\begin{tabular}{lrrrrrr}
\toprule
Method & \textsc{ti} solved & \textsc{tr} tried/solved & Final & Cost/pass & Cost/pass@\textsc{ti} & Evals/pass \\
\midrule
\system{} without GM & $25.0\!\pm\!1.7$ & $8.0\!\pm\!1.7/2.7\!\pm\!1.5$ & $27.7\!\pm\!1.2$ & $\mathbf{2.6\!\pm\!0.4}$ & $\mathbf{1.9\!\pm\!0.4}$ & $11.7\!\pm\!1.1$ \\
Full \system{} & $\mathbf{27.0\!\pm\!1.0}$ & $6.0\!\pm\!1.0/\mathbf{4.0\!\pm\!1.7}$ & $\mathbf{31.0\!\pm\!1.0}$ & $2.8\!\pm\!0.3$ & $2.4\!\pm\!0.3$ & $\mathbf{8.1\!\pm\!0.8}$ \\
\midrule
Full \system{} + Long-\textsc{ti} & $27.0\!\pm\!1.0$ & $6.0\!\pm\!1.0/1.7\!\pm\!1.5$ & $28.7\!\pm\!1.5$ & $3.2\!\pm\!0.5$ & $2.4\!\pm\!0.3$ & $8.0\!\pm\!0.9$ \\
\bottomrule
\end{tabular}
\end{table*}

\subsection{Component Analysis (RQ2--RQ3)}
\label{subsec:components}

To attribute \system{}'s repair coverage to its components, we use a matched
frozen cohort of 33 failed candidates, holding the model, structured verifier
evidence, retry policy, budgets, stopping rules, and native verifier fixed.
We call this cohort \emph{frozen-33}; Table~\ref{tab:components} reports its results.
The cohort is the 33 tasks that one shared Generator stage failed (it passed the other 59).
The cohort is chosen after the fact: fixing the starting candidates removes Generator
variance, but the study then measures repair only on tasks the Generator found hard. On this
cohort we ask three questions. First, we compare \system{}'s
single-attempt repair against the ACE-RTL baseline
(Section~\ref{sssec:single-attempt}). Second, we ask whether a fresh restart (\textsc{tr})
outperforms spending the same budget on a longer \textsc{ti}
(Section~\ref{sssec:fresh-restart}). This variant, Long-\textsc{ti}, continues \textsc{ti}
for the combined \textsc{ti} and \textsc{tr} budget without a restart. Third, we measure how
cross-task global memory (GM) changes repair coverage, verifier use, and cost
(Section~\ref{sssec:gm-eval}).

\subsubsection{Single-Attempt Comparison}
\label{sssec:single-attempt}
Table~\ref{tab:ace-single-trajectory} compares the systems when each receives
one repair attempt.
Without a fresh \textsc{tr}, the three \system{} \textsc{ti} runs reach 87/92, 84/92, and 84/92
($85.0\pm1.7$), compared with 83/92, 82/92, and 85/92 ($83.3\pm1.5$) for ACE-RTL.
In the matched frozen-33 repair study, ACE-RTL repairs 24 tasks (83/92
overall), while \system{} \textsc{ti} repairs 27 (86/92 overall) at lower model cost and
with fewer evaluations.

\begin{table}[t]
\caption{CVDP single-attempt comparison. Panel~1: fresh full-92 runs of both systems,
with \system{} keeping only \textsc{ti} ($n=3$); panel~2: matched frozen-33 repair on the shared
59-pass Generator stage ($n=1$). Bold marks the better value per panel.}
\label{tab:ace-single-trajectory}
\centering
\footnotesize
\setlength{\tabcolsep}{3pt}
\resizebox{\columnwidth}{!}{%
\begin{tabular}{l@{\hspace{2pt}}rrrrr}
\toprule
\multicolumn{6}{l}{\textit{Full-92 end-to-end, one repair attempt}} \\
\multicolumn{3}{l}{System} & Final (/92) & Cost/pass & Evals/pass \\
\midrule
\multicolumn{3}{l}{ACE-RTL serial} & $83.3\!\pm\!1.5$ & $1.5\!\pm\!0.2$ & $5.0\!\pm\!0.6$ \\
\multicolumn{3}{l}{\system{} \textsc{ti}-only} & $\mathbf{85.0\!\pm\!1.7}$ & $\mathbf{1.0\!\pm\!0.1}$ & $\mathbf{4.7\!\pm\!0.5}$ \\
\midrule
\multicolumn{6}{l}{\textit{Matched frozen-33 repair}} \\
System & Shared initial & Repairs (/33) & Final (/92) & Cost/pass & Evals/pass \\
\midrule
ACE-RTL serial & 59 & 24 & 83 & 1.3 & 4.9 \\
\system{} \textsc{ti} & 59 & \textbf{27} & \textbf{86} & \textbf{0.9} & \textbf{4.4} \\
\bottomrule
\end{tabular}}
\end{table}

\subsubsection{Fresh Restart}
\label{sssec:fresh-restart}
Fresh \textsc{tr} adds on average 2.7 repairs for \system{} without GM and 4.0 for full
\system{} (3.4 per campaign across six). Under an equal nominal
budget, fresh \textsc{tr} solves 12 of 18 matched cases in which \textsc{ti} failed, compared with
5 for Long-\textsc{ti}. Eight cases are solved only by fresh \textsc{tr}, and one only by Long-\textsc{ti} (exact McNemar test, $p=0.039$).
The fresh restart is more effective overall, although the two strategies solve
some different tasks.
In all eight \textsc{tr}-only passes, the passing design rewrites
70--175\% of the failed \textsc{ti} incumbent's lines: a failed \textsc{ti}
refines one flawed design, whereas \textsc{tr} samples a new one.

\subsubsection{Global Memory}
\label{sssec:gm-eval}
A full 92-task \system{} variant without GM reaches 86/92 (62 initial passes and
24 repairs), but its initial candidates differ from those of the main runs, so
that comparison cannot attribute the difference to memory. We therefore measure
the GM effect on the frozen-33 cohort. The three GM and no-GM repetitions keep all these matched settings fixed. GM raises mean repair
coverage from $27.7\pm1.2$ to $31.0\pm1.0$ tasks ($+3.3$, 95\% CI $[1.0,6.3]$; better on
6 tasks, worse on none, sign test $p=0.031$) and reduces mean verifier
evaluations from $324.3\pm34.1$ to $250.7\pm23.7$ (22.7\%). Mean model cost
increases from $\$70.9$ to $\$86.5$ (22.0\%), so fewer verifier calls do not
imply lower model cost.
The cost \emph{per solved task}, however, grows only from \$2.6 to \$2.8 (8.8\%) because GM solves more tasks; verifier evaluations per solved task even fall, from $11.7$ to $8.1$.
The three tasks that only the GM condition solves cite a cross-task record in every
passing run. Compared repetition by repetition, however, 2 of the 10 passes obtained only with GM cite no record, so part
of the gain is run-to-run variance. In an illustrative single run on the same frozen
cohort (Panel~2 of Table~\ref{tab:cvdp-main}), full \system{} repairs 32 of the
33 tasks (91/92 overall), compared with 24 for ACE-RTL.
Of the 56 records cited, 34 describe attempts that did not improve the verifier rank and 22
improving ones: agents cite records of failed attempts at least as often as successful ones,
consistent with fewer verifier calls. Model cost rises because every turn reads
the index.
As designed (Section~\ref{subsec:gm}), GM is shared
within one benchmark campaign, and the ablation measures exactly this setting.
Transfer to unrelated designs is not tested, and with $n{=}3$ the estimate is uncertain.

\subsection{Extension: Other CVDP Splits}
\label{subsec:heavy}
\begin{table}[t]
\caption{Other CVDP splits (Kimi K3, no baseline, $n=1$). Non-agentic and agentic-heavy are
CVDP splits; only the latter uses the LLM investigator (Section~\ref{subsec:heavy}). cidNNN: CVDP category ID. Draft~1: passed by the first
evaluated draft.}
\label{tab:heavy}
\centering
\footnotesize
\setlength{\tabcolsep}{3pt}
\begin{tabular*}{\columnwidth}{@{\extracolsep{\fill}}lrrrrr@{}}
\toprule
Split / category & Tasks & Draft~1 & Final & \$/pass & Evals/pass \\
\midrule
\multicolumn{6}{@{}l}{\emph{Non-agentic, without investigator}} \\
\quad Code completion (cid002) & 94 & 52 & 86 & & \\
\quad Spec-to-RTL (cid003) & 78 & 45 & 77 & & \\
\quad Code modification (cid004) & 55 & 32 & 49 & & \\
\quad Code improvement (cid007) & 40 & 30 & 39 & & \\
\quad Debugging (cid016) & 35 & 23 & 35 & & \\
\quad Total & 302 & 182 & \textbf{286} & 1.04 & 4.6 \\
\midrule
\multicolumn{6}{@{}l}{\emph{Agentic-heavy, with investigator}} \\
\quad Debugging (cid016) & 56 & 29 & 54 & & \\
\quad Code modification (cid004) & 6 & 3 & 6 & & \\
\quad Reuse/integration (cid005) & 7 & 1 & 7 & & \\
\quad Spec-to-RTL (cid003) & 6 & 3 & 6 & & \\
\quad Total & 75 & 36 & \textbf{73} & 1.74 & 4.6 \\
\bottomrule
\end{tabular*}
\end{table}
On CVDP's 302-task non-agentic code-generation split, with the same model,
policy, and budget as the main CVDP runs, a single run passes 286 of 302 tasks, 182 of them on the
first draft (Table~\ref{tab:heavy}).

The agentic-heavy split has 75 tasks inside public RTL repositories (median 52
RTL files, about 17K lines) and gives no list of files to edit. Here an LLM
\emph{investigator} replaces the hand-written extraction of verifier evidence. Before each
repair, it spends up to six tool steps on the verifier's logs, waveforms, and
mismatch reports and writes findings for the Repair Agent. Probe functions it
writes are shared across tasks. Each repair episode runs as a tool-calling agent with up to 40 read-only tool calls, a cost cap of \$25
per attempt, an 1800\,s timeout, and three concurrent tasks. It passes 73 of
75 tasks (\$1.74/pass): 36 on the first draft, 15 more within the five drafts,
and 22 after repair; the two failures stagnate at pass ratios of 0.67 and 0.5.

\section{End-to-End Silicon Optimization (RQ4)}
\label{sec:silicon}

All experiments so far stopped at functional correctness: a task was solved once a candidate passed the benchmark's simulation harness. Here we ask whether the same verifier-controlled acceptance extends to physical implementation. Only the verifier changes, not the method: the single simulation harness is extended with a deterministic stack of EDA and formal tools---synthesis, bounded model checking, place-and-route, timing, and energy analysis. A candidate is accepted only when these tools, never the model, report an improvement. The objective thus moves from binary pass/fail to multi-objective optimization over measured area, timing, and energy, and a bounded formal gate proves protocol properties that sampled tests can miss.
Functional correctness is only the first requirement of hardware design: a circuit that passes simulation may still be large, slow, power-hungry, or subtly violate its protocol. The case study below shows that the architecture is not specific to bug-fixing: the same controller, acceptance rule, and audit trail apply unchanged from functional repair to a full RTL-to-layout flow.
\subsection{Setting}

\subsubsection{Target and Scope}
The target is the GCD/RSA accelerator of the public CVDP task GCD 0007~\cite{doubleai_blog, qwen_blog}, which checks the key via GCD and then performs modular
exponentiation.\footnote{Its naive implementation combines a Stein GCD checker, a
square-and-multiply modular exponentiator, a shift/add multiplier, and a
combinational divider.} This is a feasibility case study rather than a
competitive benchmark: we compare an optimized design against its own naive
baseline, since none of the agentic systems of Section~\ref{sec:eval} performs
physical optimization. All gains are relative to this naive baseline, not to an expert design
or the cited industrial flows. The controller from Section~\ref{sec:architecture} serves as the repair
component of an end-to-end pipeline in which models propose RTL changes and
simulators, synthesizers, formal engines, and place-and-route tools decide
acceptance. The pipeline first runs CVDP's verifier and then continues through
physical-design measurement.

\subsubsection{Generation}
The Generator Agent from Section~\ref{subsec:repair} (Kimi K3) produces
five initial RTL candidates, each evaluated by CVDP's official Docker harness.\footnote{A locally built \texttt{nvidia/cvdp-sim:v1.0.0} image,
constructed from \texttt{cvdp\_benchmark}'s \texttt{Dockerfile.sim}, runs the
public cocotb harness (\texttt{test\_runner.py}, WIDTH 4/6/8/16, and a seeded
\texttt{COCOTB\_RANDOM\_SEED}) in a container, and a timed-out run terminates its
container by name.} The unmodified public cocotb harness drives the design with
transactions at four bit-widths (WIDTH 4/6/8/16) and checks every answer (key
validity and ciphertext) bit-exactly. The benchmark task provides six GCD module variants (three implementing Stein's algorithm), a multiplier, and an exponentiator with two swapped operands; the agent must build
the top-level controller, select the correct GCD trio, and repair the
exponentiator. Generation succeeded without any repair (\$0.43 model cost): the
winning draft instantiated the Stein trio and rewrote the exponentiator with the
correct operands, leaving every other shipped module byte-identical, and passed
all five harness tests when re-verified independently in the container (a broken design fails all five).

\subsubsection{Evaluation Stages}
Each draft goes through the tiers in Table~\ref{tab:tiers} in order and stops at the first failure. The tiers are ordered by cost:
cheap checks screen every draft, the formal tier (T1F) is the last gate before acceptance, and
the expensive physical tiers only measure accepted candidates (\emph{champions}). The harness alone
samples a few hand-written timings, checks answers but not protocol, and measures no physical properties; the remaining tiers add these checks.

\begin{table}[t]
\caption{Evaluation stages. T0--T2S screen drafts, T1F is the last gate before acceptance,
T3--T5 measure champions. txn: transaction; WNS: worst negative slack.}
\label{tab:tiers}
\centering
\footnotesize
\setlength{\tabcolsep}{3pt}
\begin{tabular}{@{}lp{0.80\columnwidth}@{}}
\toprule
Tier & Measurement \\
\midrule
T0/T1 & Icarus lint; CVDP Docker harness (cocotb, WIDTH 4/6/8/16) \\
T1R & host cocotb regression with measured cycles/txn \\
T2 & Yosys 0.46 $\rightarrow$ Nangate45 in the pinned LibreLane image: cells, area \\
T2S & OpenSTA pre-layout worst-path delay \\
T1F & SymbiYosys BMC against the flattened golden baseline (bitwuzla); UNKNOWN never passes \\
T3 & OpenROAD P\&R on the T2 netlist: die, WNS, wirelength, route DRC, DEF/GDS/SPEF timing \\
T4 & Energy/txn: OpenSTA power on the SPEF-annotated netlist $\times$ measured gate-level cycles \\
T5 & period sweep (each step $\times 0.85$): achieved clock period \\
\bottomrule
\end{tabular}
\end{table}

\subsubsection{Formal Verification}
T1F proves protocol obligations and result equivalence against the golden baseline (the generated baseline of Table~\ref{tab:gcd-rsa-e2e}) over every legal input sequence up to a bounded depth (twice the worst measured latency plus eight). It caught a bug simulation missed: in its completion state the generated FSM waits for its start input to fall before returning to idle, so a new one-cycle start pulse in that window is dropped and the done flags stay high. The harness never creates this timing, because its tests leave multi-cycle gaps between transactions. A twelve-step counterexample led to an eight-line repair (restart on the pulse) that passes both the harness and the gate.

\subsubsection{Optimization Loop}
Each round, one optimization strategy proposes a single
design change, so that each measured gain can be traced to one change (e.g., replace the combinational divider with shift/subtract
reduction); the Generator and Repair Agents modify the RTL; every draft is re-verified
through the tiers. Acceptance requires all gates passed and a strictly better
measured rank over mapped area (T2) and estimated runtime (cycles/txn $\times$
pre-layout delay) following Equation~\eqref{eq:promotion}; regressions are reverted.

\subsection{Experimental Results}

\begin{table}[t]
\caption{Campaign of the Generator and Repair Agents on the generated baseline (Kimi K3); $\Delta$ is
relative to this naive baseline, not to an expert design.}%
\label{tab:gcd-rsa-e2e}
\centering
\footnotesize
\setlength{\tabcolsep}{4pt}
\begin{tabular}{lrrr}
\toprule
Metric & Baseline & Champion & $\Delta$ \\
\midrule
Mapped area (\textmu m$^2$, T2) & 6,867.1 & 2,223.2 & $-67.6\%$ \\
Cells (T2) & 5,108 & 1,406 & $-72.5\%$ \\
Cycles/txn (T1R) & 229 & 59.6 & $-74.0\%$ \\
Die area (\textmu m$^2$, T3) & 18,231.8 & 6,170.1 & $-66.2\%$ \\
WNS at 2.0 ns (ns, T3) & $-13.69$ & $+0.51$ & $+14.20$ ns \\
Energy/txn (nJ, T4) & 0.785 & 0.130 & $-83.4\%$ \\
Achieved period (ns, T5) & 20.35 & 1.47 & $-92.8\%$ \\
\bottomrule
\end{tabular}
\end{table}

Table~\ref{tab:gcd-rsa-e2e} compares the generated baseline with the campaign's best accepted candidate (119 sessions, 1,492 evaluations, \$55 cost). The champion results from a sequence of accepted changes: the combinational divider becomes an iterative shift/subtract reduction, control logic is pruned, one subtractor is shared for compare and subtract, and an early exit skips multiply rounds when the base is one; each step was accepted only on a measured rank improvement. Mapped area falls 67.6\% (6,867.1 $\to$ 2,223.2~\textmu m$^2$), latency 74.0\% (229 $\to$ 59.6 cycles), and die area 66.2\% (18,231.8 $\to$ 6,170.1~\textmu m$^2$); at a 2.0~ns clock, timing moves from a 13.69~ns violation to 0.51~ns slack, and energy per transaction falls 83.4\% (0.785 $\to$ 0.130nJ). The champion passes the harness, regression, and formal gate, with zero route design-rule violations. Its achieved period (T5) is 1.47~ns ($\approx$680~MHz), against the baseline's 20.35~ns ($\approx$49~MHz). An earlier campaign from a hand-reconstructed baseline reached a similar reduction (die area $-61\%$).
The run stopped on its evaluation budget with 0.51~ns of slack left at
the 2.0~ns layout clock; the loop accepts candidates on mapped area and
estimated runtime, not routed timing, so this slack is unused. It could be traded for smaller,
lower-power cells, and the empty part of the fixed floorplan (below) for a
smaller die.

\begin{figure}[t]
\centering
\includegraphics[width=\columnwidth]{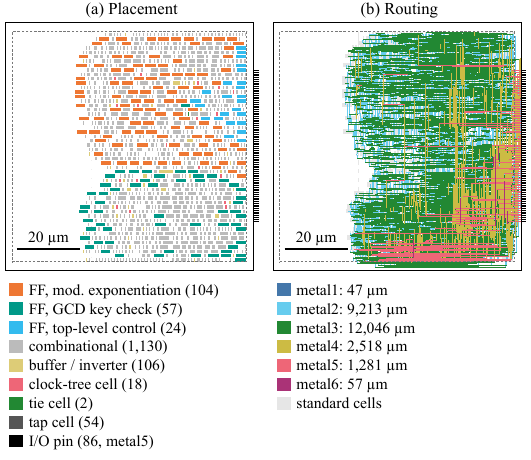}
\caption{Routed layout of the champion (Nangate45, OpenROAD, 2.0~ns
clock), drawn from the T3 DEF; every cell, wire and via matches the
streamed-out GDS. Die $78.55\times78.55$~\textmu m (solid), core
$74.29\times72.8$~\textmu m (dashed). (a)~Standard cells at their LEF sizes,
coloured by function; flip-flops (FF) are coloured by the RTL module of the
register they implement. (b)~Signal and clock routing on metal1--metal6 at
drawn wire widths, with wirelength per layer. All shapes are to scale except
the I/O-pin marks on the east edge, which are enlarged.}
\label{fig:layout}
\end{figure}

Fig.~\ref{fig:layout} shows the champion's routed layout. The flow
sizes every candidate's die in the same way: a square core at 40\% utilization
of the mapped area, with cells placed at a global-placement target density of
0.6, so die area follows mapped area. After timing repair, clock-tree synthesis
and tap insertion, the 1,495 cells (1,441 instances and 54 tap cells,
2,699~\textmu m$^2$) occupy 49.9\% of the core. The 185 flip-flops group by
module: modular exponentiation (104) mostly in the north, GCD key check (57)
mostly in the south, and top-level control (24) along the east edge, where the
pin placer put all 86 I/O pins. Because the placer spreads cells only down to
the target density and wirelength pulls them toward the pins, a cell- and
wire-free strip forms on the west side and covers 31\% of the core. It is
unused area of the fixed floorplan, not a placement defect. Routing totals
25.2~mm on metal1--metal6 (85\% on metal2 and metal3), against 110.4~mm on all
ten metal layers for the baseline; clock-tree synthesis adds 17 clock buffers
and one load cell.

\section{Beyond the Official Testbenches (RQ5)}
\label{sec:formal}
A task counts as solved when a design passes the benchmark's testbench (on
CVDP, a hidden cocotb harness). Since a testbench samples the input space, a
passing design can still be wrong. We measure how often this happens for our
designs and turn the wrong ones into tests that reject them.

\subsection{Designs and References}
We study the 448 passing CVDP designs of our Kimi K3 runs: 89 agentic designs
from \system{}, and 286 non-agentic and 73 agentic-heavy designs from
Section~\ref{subsec:heavy}. A second agentic run produced passing designs for 23 of the same tasks, and
ChipBench contributes 39 passing designs from an earlier run. CVDP ships no
golden designs, so the reference depends on the task category, giving four studies: (A)~for 39
code-improvement tasks, the original RTL, which the design must keep
equivalent; (B)~for 46 heavy bug-fix tasks, the upstream file into which CVDP
injected the bug; (C)~for the 23 tasks with two passing designs, the other
design; and (D)~for 119 modification and debug tasks, the original RTL with the
requested change's behaviour excluded from the comparison by input constraints or a wrapper. ChipBench ships a
reference design per task.

\subsection{Checking Method}
A design and its reference receive the same inputs, and bounded model checking
searches for an output difference within 20 cycles after reset (yosys with
slang, SymbiYosys, bitwuzla). A difference must survive a four-valued recheck,
which discards effects of uninitialized registers, and must be reproduced in
simulation; a no-difference result counts only if the outputs change within the bound, so that the check is not vacuous. For 34 designs of study~D the check cannot run directly (e.g., two clock
domains, an original that does not compile, real arithmetic); they are checked through a wrapper or a minimally repaired original, by
showing that the original modules are unchanged, or in simulation on legal
inputs. We then read
each difference against the specification: it counts as a wrong design only on
allowed inputs, at a time the output is defined. Two disputed differences are
not counted.

Two studies need no reference. (E)~CVDP runs each harness once; we rerun the 399
designs whose harness is simulation-based, each under 40 seeds, and examine every failure and
every mismatch the harness only logs. (F)~To measure how many errors the harnesses catch, up to ten
single-change mutants of each of 277 designs run through the design's harness,
and, for the 189 designs the formal check handles, every accepted mutant is
compared formally with its design, separating missed mutants from equivalent
ones. ChipBench is measured with 438 mutants of
37 references.

\begin{table}[t]
\caption{Designs that pass the official testbench but are functionally wrong.
With ref.: checked against a reference or a second design; the others were at
most rerun with other seeds~(E), so counts are lower bounds. Second-run designs
are not among the 448. Mutation scores exclude equivalent mutants (up to 10
mutants per CVDP design, up to 40 per ChipBench reference).}
\label{tab:formal}
\centering
\footnotesize
\setlength{\tabcolsep}{3pt}
\resizebox{\columnwidth}{!}{%
\begin{tabular}{@{}llrrr@{}}
\toprule
Benchmark & Reference (studies) & Passed & With ref. & Wrong \\
\midrule
CVDP agentic (92) & second design, original (C, D) & 89 & 47 & 4 \\
\quad second run (shared tasks) & \system{} design (C) & 23 & 23 & 3 \\
CVDP non-agentic (302) & original (A, D); seeds (E) & 286 & 123 & 14$^{a}$ \\
CVDP heavy (75) & upstream file (B) & 73 & 46 & 13$^{b}$ \\
\midrule
CVDP total & & 448\,+\,23 & 216\,+\,23 & \textbf{31\,+\,3} \\
ChipBench (43) & reference module & 39 & 39 & 1 \\
\midrule
\multicolumn{5}{@{}l}{\emph{Formal mutation score of the testbenches}} \\
CVDP harnesses & 189 designs, 1,282 differing mutants & \multicolumn{3}{r}{74\% rejected} \\
ChipBench testbenches & 37 references, 410 differing mutants & \multicolumn{3}{r}{93\% rejected} \\
\bottomrule
\multicolumn{5}{@{}l}{$^{a}$11 found against a reference, 3 by the seed sweep (one of moderate confidence).} \\
\multicolumn{5}{@{}l}{$^{b}$One affects performance only (spurious pipeline flushes).}
\end{tabular}}
\end{table}

\subsection{Findings}
At least 31 of the 448 CVDP designs are functionally wrong, as are 3 of the 23
second-run designs (Table~\ref{tab:formal}). Against references A--C, 21 of the
71 comparisons with a definite verdict reveal a wrong design. One heavy fix
leaves the injected bug in place, so a UDP receiver flags every valid packet as
an error; in another, a CHERIoT Ibex core traps on every store whose offset
has bit~4 set. The seed sweep finds three more; a stopwatch fails on 21 of 40 seeds and
passed CVDP's single run by chance.

The CVDP harnesses reject only 74\% of the mutants that differ from their design,
against 93\% for ChipBench. This is not an artifact of illegal inputs: 39 of 40 sampled accepted differing
mutants differ on legal inputs. Six harnesses cannot fail; one
compares its Python model with itself. ChipBench's testbenches are uneven: they
reject only 35\% of a MIPS coprocessor-register task's mutants and accept a
sequence detector that ignores the first bit after reset. The CVDP harnesses
miss errors because inputs are held for several cycles, each test covers one
configuration, random tests draw too few samples, or a check only logs a
mismatch.

\subsection{Strengthened Testbenches}
For each wrong design we write a short input sequence that follows the
specification and exposes the error. Simulating the reference on it with the
harness's simulator yields the expected values, kept only where the
specification defines the output; none is written by hand. The resulting cocotb
test is added to the task's harness in CVDP's Docker image and is confirmed when
the reference passes every test and the wrong design passes the original tests
but fails the added one. Where the original RTL lacks the new feature or still
has the bug, or no original exists (seed-sweep tasks), the reference is the
wrong design with only the bug fixed.

Tests are confirmed for 25 of the 34 wrong CVDP designs (3 to 39 cycles). Of the
other nine, five need what the single-clock flow lacks (two clock domains, a
gated clock, a real-valued table, or an unbuilt configuration), two need
stimulus the harness cannot apply (about 90,000 cycles, or an input its wrapper cannot
drive), and in two the harness itself is faulty. For ChipBench, a task's
counterexamples form an additional testbench with the official clock and
sampling; it rejects the wrong sequence detector and all 27 missed mutants. When built from either half of the coprocessor task's missed mutants, it rejects 12 of the 14 and all 12 mutants of the other half. We package 14 CVDP tests, which drive only inputs
the original harness uses, and both ChipBench testbenches as an additional evaluation, hidden from the agents,
beside the unchanged official score. The other 11 confirmed CVDP tests belong
to heavy tasks outside our packaged dataset (9) or would fix a latency the
specification leaves open (2).

\begin{table}[t]
\caption{Runs re-scored with the added tests (Kimi K3, mean $\pm$ std., $n=3$;
frozen-33 and Section~\ref{subsec:heavy}: $n=1$). inv.: LLM investigator.
Strengthened: a task also needs the added test.}
\label{tab:strict}
\centering
\footnotesize
\setlength{\tabcolsep}{4pt}
\begin{tabular}{@{}llrr@{}}
\toprule
Benchmark & System & Official & Strengthened \\
\midrule
CVDP (/92) & \system{} & $88.7\pm0.6$ & $87.3\pm0.6$ \\
 & ACE-RTL serial & $83.3\pm1.5$ & $82.7\pm2.1$ \\
 & MAGE local & $80.7\pm4.2$ & $79.7\pm4.2$ \\
CVDP frozen-33 (/92) & \system{} / ACE-RTL & 91 / 83 & 89 / 82 \\
ChipBench (/43) & all systems & \multicolumn{2}{c}{unchanged} \\
\midrule
CVDP non-agentic (/302) & \system{} & 286 & 275 \\
CVDP heavy (/75) & \system{} w/ inv. & 73 & 64 \\
\bottomrule
\end{tabular}
\end{table}

\subsection{Effect on the Reported Results}
We re-score all runs with these 14 CVDP tests (3 agentic, 11
non-agentic), the 9 heavy tests and both ChipBench testbenches
(Table~\ref{tab:strict}). The tests come from our own runs, two of them with a
\system{} design as reference, so they target our agents' errors and may miss errors that only the baseline systems make. With the added tests, \system{} solves 1.3 fewer CVDP tasks per run on
average (4 fewer over its three runs), ACE-RTL 0.7 fewer (2) and MAGE 1.0 fewer
(3); the ranking is unchanged. No ChipBench score changes: no run solves
the coprocessor task, and the wrong sequence detector comes from this study's
earlier run. The Section~\ref{subsec:heavy} runs fall from 286 to 275 and from
73 to 64, by construction, since the tests were built from their wrong designs.
The check is bounded and single-clock, and the classification rests on our
reading of each specification, so the counts are lower bounds for our runs, not
estimates for other systems. The main results keep the official protocol for
comparability with published scores.
\section{Conclusions}
We presented ChipArc, a multi-agent system that separates RTL generation from acceptance: models propose changes, and a deterministic controller checks every candidate with the configured verifier before updating the accepted design. Because structured repair, bounded exploratory branches, and fresh restarts share one acceptance rule, no unverified candidate can displace a better verified one. On CVDP, ChipArc achieves the highest completion at the lowest model cost among the evaluated systems while using a verifier budget comparable to serial ACE-RTL; matched studies favor fresh restarts after stagnation and show that cross-task memory improves repair coverage and reduces verifier calls at higher model cost. With the same acceptance rule extended to physical design, \system{} improves the generated GCD/RSA baseline by 68\% in mapped area, 74\% in latency, and 83\% in energy per transaction, with layouts closing at 1.47ns ($\approx$680MHz) within \$55 model cost, and its formal check exposed a protocol bug that simulation missed. Checking passing CVDP and ChipBench designs against reference designs
exposed wrong designs that the official testbenches accept. We turned the
counterexamples into tests that reject 25 of the 34 wrong CVDP designs and the
wrong ChipBench design, and package 14 CVDP tests and both ChipBench testbenches
as an optional stricter evaluation.
Audit logs bind every accepted transition to a recorded verifier outcome, supporting model-free offline replay of acceptance decisions. Our guarantees are only as strong as the verifiers,
which on CVDP and ChipBench are simulation tests. Baselines use modified,
non-compute-matched configurations; memory is evaluated only on batches of
related tasks; physical gains are relative to a naive baseline. Requiring the strengthened tests (Section~\ref{sec:formal}) costs each compared system about one
CVDP task on average and leaves the ranking unchanged.%

\bibliographystyle{IEEEtran}

\end{document}